\documentclass[prl,showpacs,twocolumn,floatfix,superscriptaddress]{revtex4}

\usepackage{times,amsmath,amsfonts,amssymb,latexsym}
\usepackage{graphicx,epsf}

\newcommand{\be}{\begin{equation}}
\newcommand{\ee}{\end{equation}}
\newcommand{\ba}{\begin{eqnarray}}
\newcommand{\ea}{\end{eqnarray}}

\newcommand{\ignore}[1]{}
\def\CC{{\rm\kern.24em \vrule width.04em height1.46ex depth-.07ex
    \kern-.30em C}}
\def\P{{\rm I\kern-.25em P}}
\def\RR{{\rm
         \vrule width.04em height1.58ex depth-.0ex
         \kern-.04em R}}

\def\bbbc{{\mathchoice {\setbox0=\hbox{$\displaystyle\rm C$}\hbox{\hbox
to0pt{\kern0.4\wd0\vrule height0.9\ht0\hss}\box0}}
{\setbox0=\hbox{$\textstyle\rm C$}\hbox{\hbox
to0pt{\kern0.4\wd0\vrule height0.9\ht0\hss}\box0}}
{\setbox0=\hbox{$\scriptstyle\rm C$}\hbox{\hbox
to0pt{\kern0.4\wd0\vrule height0.9\ht0\hss}\box0}}
{\setbox0=\hbox{$\scriptscriptstyle\rm C$}\hbox{\hbox
to0pt{\kern0.4\wd0\vrule height0.9\ht0\hss}\box0}}}}
\def\bbbz{{\mathchoice {\hbox{$\sf\textstyle Z\kern-0.4em Z$}}
{\hbox{$\sf\textstyle Z\kern-0.4em Z$}}
{\hbox{$\sf\scriptstyle Z\kern-0.3em Z$}}
{\hbox{$\sf\scriptscriptstyle Z\kern-0.2em Z$}}}}

\begin{document}

\title{Adiabatic Preparation of Topological Order}
\author{Alioscia Hamma}
\affiliation{Department of Chemistry, University of Southern California, Los Angeles, CA
90089, USA}
\author{Daniel A. Lidar}
\affiliation{Department of Chemistry, University of Southern California, Los Angeles, CA
90089, USA}
\affiliation{Departments of Electrical Engineering and Physics, University of Southern
California, Los Angeles, CA 90089, USA}

\begin{abstract}
Topological order characterizes those phases of matter that defy a
description in terms of symmetry and cannot be distinguished in terms local
order parameters. This type of order plays a key role in the theory of the
fractional quantum Hall effect, as well as in topological quantum
information processing. Here we show that a system of $n$ spins forming a
lattice on a Riemann surface can undergo a second order quantum phase
transition between a spin-polarized phase and a string-net condensed phase.
This is an example of a quantum phase transition between magnetic and
topological order. We furthermore show how to prepare the topologically
ordered phase through adiabatic evolution in a time that is upper bounded by 
$O(\sqrt{n})$. This provides a physically plausible method for constructing
and initializing a topological quantum memory. 
\end{abstract}

\maketitle

The notion of \emph{topological order} \cite{Wen3} can explain phase
transitions that depart from the standard Landau theory of symmetry
breaking, long range correlations and local order parameters. It plays a key
role in condensed matter theory of strongly correlated electrons; e.g., it
provides a means to understand the different phases arising in the
fractional quantum Hall effect \cite{Wen:90}. These phases have exactly the
same symmetries. No \emph{local} order parameter can distinguish them. The
internal order that characterizes these phases is topological, and can be
characterized in terms of a \emph{non-local} order parameter, e.g., the
expectation value of string operators (operators formed by tensor products
of local operators taken along a string). Recently, it was shown that
topological order can also be characterized by topological entropy \cite%
{Kitaev:06Levin:06}.

Another arena in which topological order has found profound applications is
quantum computation, in particular in the context of systems exhibiting
natural fault tolerance \cite{Kitaev:03,Freedman:03}. Central to this
application is the ability to prepare certain topologically ordered states,
which are the ground states of a Hamiltonian describing a spin system
occupying lattice links on a Riemann surface of genus $\mathfrak{g}$ (e.g.,
a torus, for which $\mathfrak{g}=1$). In a topologically ordered phase the
ground state degeneracy depends on $\mathfrak{g}$. For instance, in the
toric code \cite{Kitaev:03} the ground state manifold $\mathcal{L}$ is $2^{2 
\mathfrak{g}}$-fold degenerate, and this allows one to encode $2\mathfrak{g}$
qubits. Because any two orthogonal states in $\mathcal{L}$ are coupled only
by loop operators with loops that wind around the holes of the Riemann
surface, this encoding is robust against any local perturbation. Such a
system constitutes a very robust memory register \cite{Kitaev:03}.

Here we address the problem of \emph{preparation of topologically ordered\
states via\ adiabatic evolution}. Our motivation for studying this problem
is at least threefold. \emph{First}, while quantum phase transitions (QPTs)
between deconfined and confined phases have been studied extensively in
quantum chromodynamics \cite{Wilson:74}, this has not been the case in
condensed matter physics. Here we study a second order QPT between a
magnetically ordered phase (deconfined, topologically disordered) and a
string-net condensed phase \cite{Levin:05} (confined, topologically
ordered). \emph{Second}, we are motivated by the aforementioned need for
topologically ordered states as the input to topological quantum computers.
In Ref.~\cite{Dennis:02} it was shown how to prepare a ground state of the
toric code via $O(\sqrt{n})$ repeated syndrome measurements, where $n\sim
L^{2}$ is the number of spins on a lattice of linear dimension $L$. Such
ground state preparation is an essential step in constructing a
topologically fault tolerant quantum memory. However, preparation via
syndrome measurements has certain drawbacks, most notably that the
measurements are assumed to be as fast as logic gates and must be fast
compared to the decoherence timescale, assumptions which are likely to be
hard to satisfy in practice. A\ system in which the phase transition
predicted here can potentially be realized experimentally, as well as used
for topological quantum memory, is a Josephson junction array \cite{Ioffe:02}%
. \emph{Third}, we believe that the methods presented here will also find
applications in the field of adiabatic quantum computing \cite{Farhi:01},
for the preparation time in our adiabatic method is $O(\sqrt{n})$ (as in the
syndrome measurement method \cite{Dennis:02}), which is optimal in the sense
that it saturates the Lieb-Robinson bound \cite{Bravyi:06Eisert:06}.

\textit{The toric code model}.--- Consider an (irregular) lattice on a
Riemann surface of genus $\mathfrak{g}$. At every link of the lattice we
place a spin $1/2$. If $n$ is the number of links, the Hilbert space of such
a system is $\mathcal{H}=\mathcal{H}_{1}^{\otimes n}$, where $\mathcal{H}%
_{1}=\mathrm{Span}\{|0\rangle ,|1\rangle \}$ is the Hilbert space of a
single spin. In the usual computational basis we define a reference basis
vector $|0\rangle \equiv |0\rangle _{1}\otimes ...\otimes |0\rangle _{n}$,
i.e., all spins up. This string-free state is the vacuum state for strings.
We define the Abelian group\ $X$ as the group of all spin flips on $\mathcal{%
H}$. Its elements can be represented in terms of the standard Pauli matrices
as $x_{\alpha }=(\sigma _{1}^{x})^{\alpha _{1}}\otimes ...\otimes (\sigma
_{n}^{x})^{\alpha _{n}}$, where $\alpha _{j}\in \{0,1\}$ and $\alpha
=\{\alpha _{j}\}_{j=1}^{n}$. Every $x_{\alpha }\in X$ squares to identity $%
\mathbb{I}$, $\dim {\mathcal{H}}=|X|=2^{n}$, and any computational basis
vector can be written as $|\alpha \rangle =x_{\alpha }|0\rangle $.

\begin{figure}[tbp]
\begin{equation*}
\leavevmode\hbox{\epsfxsize=4.5 cm 
   \epsffile{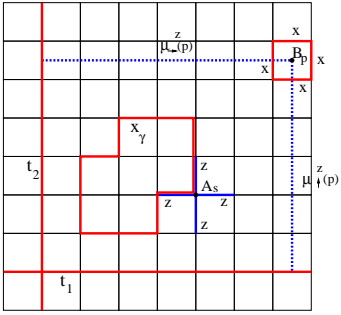}}
\end{equation*}%
\caption{(color online) The square lattice on a torus. Closed $x$-strings
(in red) commute with the star operator $A_{s}$ because they have an even
number of links in common. The elementary closed $x$-string is the plaquette 
$B_{p}$. Dashed $z$-strings connect plaquettes $p$. The two incontractible $%
x $-strings are denoted $t_{1,2}$. The dual string variable (dashed, blue)
connects the reference lines $t_{1},t_{2}$ with the vertices $p$ in the dual
lattice. Thus the dual operator $\protect\mu _{\rightarrow (\uparrow
)}^{z}(p)$ anti-commutes with the plaquette variable $\protect\mu %
^{x}(p)\equiv B_{p}$.}
\label{lattice}
\end{figure}

Now we associate a geometric interpretation with the elements of $X$, as in
Fig.~\ref{lattice}. 
To every string $\gamma $ connecting any two vertices of the lattice we can
associate the string operator $x_{\gamma}$ operating with $\sigma ^{x}$ on
all the spins covered by $\gamma $. The product of two string operators is a 
\emph{string-net} operator: $x_{\gamma }\equiv x_{\gamma _{1}}x_{\gamma
_{2}} $, and we can view $X$ as the group of string-net operators on the
lattice. A particularly interesting subgroup of $X$ is formed by the
products of all \emph{closed} strings; we denote this \emph{closed
string-nets} group \cite{Wen3} by $\overline{X}$. If the Riemann surface has
genus $\mathfrak{g}$, we can also draw $2\mathfrak{g}$ incontractible
strings around the holes in the surface: $T=\langle t_{1},...,t_{2g}\rangle $%
; see Fig.~\ref{lattice}. The group of contractible closed strings is
denoted by $\mathcal{B}$, and is generated by the elementary closed strings: 
$B_{p}=\prod_{j\in \partial p}\sigma _{j}^{x}$, where $j$ labels all the
spins located on the boundary $\partial p$ of a plaquette $p$. A \emph{star}
operator is defined as acting with $\sigma ^{z}$ on all the spins coming out
of the vertex $s$: $A_{s}=\prod_{j\in s}\sigma _{j}^{z}$. Then $\overline{X}$
splits into 
$2\mathfrak{g}$ topological sectors labeled by the $t_i$: $\overline{X}
=T\cdot \mathcal{B}$ \cite{Hamma:05a}.

\textit{Topological order in the toric code model}.--- Consider the
Hamiltonian $H(g,U)=-g\sum_{p}B_{p}-U\sum_{s}A_{s}$. For $g=U=1$ this is the
Hamiltonian introduced by Kitaev \cite{Kitaev:03}, and we will refer to it
as the Kitaev Hamiltonian. It is an example of a $Z_{2}$-spin liquid, a
model that features string condensation in the ground state. The operators $%
A_{s},B_{p}$ all commute, so the model is exactly solvable, and the ground
state manifold is given by $\mathcal{L}=\{|\phi \rangle \in \mathcal{H\;}%
|\;A_{s}|\phi \rangle =B_{p}|\phi \rangle =|\phi \rangle \}$. The vector in
the trivial topological sector of $\mathcal{L}$ is explicitly given by \cite%
{Hamma:05a} 
\begin{equation}
|\phi _{0}\rangle =|\mathcal{B}|^{-\frac{1}{2}}\sum_{x\in \mathcal{B}
}x|0\rangle .  \label{eq:phi0}
\end{equation}
On a torus the ground state manifold $\mathcal{L}$ is $4$-fold degenerate. A
vector in $\mathcal{L}$ can be written as the superposition of the four
orthogonal ground states in different topological sectors $|\phi _{k}\rangle
=t_{1}^{i}t_{2}^{j}|\phi _{0}\rangle ,i,j\in \{0,1\},\,k=i+2j$. On an
arbitrary lattice (regular or irregular) on a generic Riemann surface of
genus $\mathfrak{g}$ there are $2\mathfrak{g}$ such operators $t_{i}$ and
thus the degeneracy of the ground state manifold is, in general $2^{2 
\mathfrak{g}}$. This is a manifestation of \emph{topological order} \cite%
{Wen3}.

\textit{A quantum phase transition}.--- Consider the Hamiltonian obtained by
applying a magnetic field in the $z$-direction to all the spins: 
\begin{equation}
H_{\xi }=-\xi \sum_{j=1}^{n}\sigma _{j}^{z}+\xi nI.
\end{equation}
The zero-energy ground state of this Hamiltonian is the magnetically
ordered, spin-polarized string-vacuum state $|0\rangle $. In the presence of 
$H_{\xi }$ a string state $x_{\gamma }|0\rangle $, with $x_{\gamma
}=\prod_{j\in \gamma }\sigma _{j}^{x}\neq \mathbb{I}$, pays an excitation
energy that depends only on the string length $l_{\gamma }$, namely\ $%
\langle 0|x_{\gamma }H_{\xi }x_{\gamma }|0\rangle =2\xi l_{\gamma }$. Thus $%
H_{\xi }$ is a \emph{tension term}. If we add the star term $%
H_{U}=-U\sum_{s}A_{s}$ to $H_{\xi }$, the ground state is still $|0\rangle $
[its energy is $E_{g}(U)=-Un_{s}$], but the spectrum is quite different. Now
there is a drastic difference between open and closed $x$-type strings.
Closed strings commute with all the $A_{s}$ and hence only pay the tension,
whereas open strings also pay an energy of $2U$ for every spin that is
flipped due to anti-commutation ($-U\langle 0|\sigma _{j}^{x}A_{s}\sigma
_{j}^{x}|0\rangle =+U$). So every open string pays a total energy of $4U$.
In the $U\gg \xi $ limit, open strings are energetically forbidden. On the
other hand, the plaquette operators $B_{p}$ deform $x$-strings: the
plaquette term $H_{g}=-g\sum_{p}B_{p}$ acts as a kinetic energy for the
(closed) strings and induces them to fluctuate. In light of these
considerations the total model $H=H_{\xi }+H(g,U)$ has two different phases.
The first phase is the \emph{spin polarized} phase described above, when $%
U\gg \xi \gg g\sim 0$. Here the string fluctuations are energetically
suppressed by the tension, so the ground state is $|0\rangle $. The other
phase arises when $U\gg g\gg \xi \sim 0$. Now the tension is too weak to
prevent the closed loops from fluctuating strongly. Nevertheless, open $x$%
-type strings cost the large energy $4U$, so are forbidden in the ground
state. The ground state consists of the superposition with equal amplitude
of all possible \emph{closed} strings, $\mathcal{L}$. This is the \emph{%
string-net condensed phase}. As the tension decreases, and the fluctuations
increase, the ground state becomes a superposition of an increasingly larger
number of closed strings. Open strings cannot be present in the ground state
because their energy is too large. In the thermodynamic limit, for some
critical value of the ratio between tension and fluctuations $\xi /g$, the
gap between the ground state and the first excited state closes and the
system undergoes a second order QPT. Notice that the gap between the ground
and first excited state is given only by the interplay between tension and
fluctuations, and is unaffected by $H_{U}$. The 
second phase 
we have described cannot be characterized by a local order parameter, such
as magnetization: we have phases without symmetry breaking. This is an
example of topological order. The low energy sector has energy much smaller
than $U$ and it thus comprises closed strings: $\mathcal{L}_{\mathrm{low}}=%
\mathrm{span}\{x|0\rangle ,\;x\in \overline{X}\}$. This subspace also splits
into four sectors labeled by the elements of $T$: $\mathcal{L}_{\mathrm{low}%
}^{ij}=\mathrm{span}\{x t_{1}^{i}t_{2}^{j} |0\rangle ,x\in \mathcal{B}%
;i,j=0,1\}$. Thus $\dim \mathcal{L}_{\mathrm{low}}=\dim (\mathcal{H}%
)/(2^{n/2+1})$. Within each sector, low level excitations have an energy $E_{%
\mathrm{low}}\leq g$ and become gapless only at the critical point in the
thermodynamic limit.

\textit{Adiabatic evolution}.--- We now show how to prepare the
topologically ordered state $|\phi _{0}\rangle $ through adiabatic
evolution. The idea is to adiabatically interpolate between an initial
Hamiltonian $H(0)$ whose ground state is easily preparable, and a final
Hamiltonian $H(1)$ whose ground state is the desired one. The interpolation
between two such Hamiltonians has been used as a paradigm for adiabatic
quantum computation \cite{Farhi:01}. This process must be accomplished such
that the error $\delta $ between the actual final state and the desired
(ideal adiabatic) ground state of $H(1)$, is as small as possible. The
problem is that real and virtual excitations can mix excited states with the
desired final state, thus lowering the fidelity \cite{SarandyLidar:05}.
Rigorous criteria are known \cite{hagedorn,Ruskai:06}, and used below, which
improve on the first-order approximation typically encountered in quantum
mechanics textbooks. For smooth interpolations and one relevant excited
state it has been proven that an arbitrarily small error $\delta $ is
obtained when the evolution time $T$ obeys $T=O(1/\Delta E_{\min })$, where $%
\Delta E_{\min }$ is the minimum energy gap encountered along the adiabatic
evolution \cite{Schaller:06}.

Consider the following time-dependent Hamiltonian: 
\begin{equation}
H(\tau )=H_{U}+\left[ 1-f(\tau )\right] H_{\xi }+f(\tau )H_{g}
\label{Htotal}
\end{equation}
Where $U\gg g,\xi \sim 1$ and $f:\tau \in \lbrack 0,1]\mapsto \lbrack 0,1]$
such that $f(0)=0$ and $f(1)=1$, $\tau =t/T$ is a dimensionless time. At $%
\tau =0$ the Hamiltonian is $H(0)=H_{U}+H_{\xi }$ whose ground state is the
spin polarized phase $|0\rangle $. At $\tau =1$ the Hamiltonian is the
Kitaev Hamiltonian $H(1)=H_{U}+H_{g}$ whose ground state is in $\mathcal{L}$%
, the string-net condensed phase. The tension strength is given by $\xi
(1-f(\tau ))$, while the fluctuations have strength $gf(\tau )$. Adiabatic
evolution from the initial state $|0\rangle $ will cause the fluctuations to
prepare the desired ground state $|\phi _{0}\rangle $. Notice that because
of the interplay between the tension term $H_{\xi }$ and the large term $%
H_{U}$, at every $\tau $ the ground state is in $\mathcal{L}_{\mathrm{low}
}^{00}$. As the tension decreases, the gap between the ground state and
states in the other sectors $\mathcal{L}_{\mathrm{low}}^{ij}$ ($i+j>0$)
diminishes. Beyond the critical point, the gap scales down exponentially
with the number of spins. Nevertheless, this does not cause transitions:
irrespective of how small the gap is, the Hamiltonian does not couple
between different sectors. Indeed, the adiabatic theorem states that the
probability of a transition between the instantaneous eigenstate $|\psi_j
(\tau )\rangle$ the system occupies and another eigenstate $|\psi_j (\tau
)\rangle$ (with respective energies $E_i(\tau),E_j(\tau)$) is given by $%
p_{ij} \le | \langle \psi_i (\tau )|\dot{H}|\psi_j (\tau )\rangle
/(E_i(\tau)-E_j(\tau))^2 |^2$. In our case, $\langle \psi_i (\tau )|\dot{H}%
|\psi_j (\tau )\rangle \equiv 0 $ whenever $|\psi_i (\tau )\rangle ,|\psi _j
(\tau )\rangle $ belong to two different sectors \cite{HLnew}. Thus
transitions between different $\mathcal{L}_{\mathrm{\ low}}$ sectors are
completely forbidden, and the adiabatic evolution keeps the ground state in
the sector it starts in. The situation is different in the presence of an 
\emph{arbitrary} $k$-local perturbation $V$: time evolution can then, in
principle, couple the different sectors, and this requires a careful
analysis. Before topological order is established different sectors remain
gapped because of the tension, and the adiabatic criterion applies. Once in
the topologically ordered phase, the exponentially small gap requires us to
apply time dependent perturbation theory to compute the probability of
tunnelling between different sectors. It turns out that only the $L/k$th
order in the Dyson series is non-vanishing, and the result is that as long $%
V<\xi$, these transitions are suppressed exponentially in $L/k$ at arbitrary
times \cite{HLnew}. Namely, the probability of transition between two
different sectors $a$ and $b$ is 
\begin{equation}  \label{ww}
W_{b\leftarrow a}(\tau )\leq |(V/\xi)^{L/k}L|^{2}\qquad \forall \tau
\end{equation}
This result is confirmed by numerical analysis \cite{numerical}: \emph{%
topological order protects the adiabatic evolution from tunneling to other
sectors}. Note that, in contrast, preparing the ground state by cooling
would result in an arbitrary superposition in the ground state manifold.
Therefore we are concerned with the gap within $\mathcal{L}_{ \mathrm{low}%
}^{00}$. We next show that this gap scales as $\Delta E(n;\tau _{c})\sim
n^{-1/2}$ at the critical point $\tau _{c}$. To do this, we must understand
the lattice gauge theory that represents the low energy sector of the model.

\textit{Lattice gauge theory and mapping to the Ising model}.--- We briefly
review the lattice gauge theory invented by Wegner \cite{Wegner:71}, showing
that the low energy sector ($E\ll U$) of Kitaev's model maps onto this
theory. The lattice gauge Hamiltonian is given by 
\begin{equation}
H_{\mathrm{gauge}}=-\lambda _{1}\sum_{s,\hat{k}}\sigma ^{z}(s,\hat{k}%
)-\lambda _{2}\sum_{p}B_{p}.  \label{3Dgauge}
\end{equation}
A local gauge transformation consists in flipping all the spins originating
from $s$, so it can be defined as: $A_{s}=\prod_{j\in s}\sigma _{j}^{z}$. It
is simple to check that $H_{\mathrm{gauge}}$ is invariant under this local
gauge transformation. Elitzur's theorem \cite{Elitzur:75} implies that the
local symmetry cannot be spontaneously broken. In a gauge theory the Hilbert
space is restricted to the gauge-invariant states. Thus the Hilbert space
for this model is given by $\mathcal{H}_{\mathrm{gauge}}=\{|\psi \rangle \in 
\mathcal{H\,} |\,\,A_{s}|\psi \rangle =|\psi \rangle \}\subset \mathcal{H}$,
the Hilbert space spanned by all possible spin configurations. The
Hamiltonian (\ref{3Dgauge}) clearly resembles our $H(\tau )$, apart from the
absence of the star term $H_{U}$. Consequently $\mathcal{H}_{\mathrm{gauge}}$
is just the low energy sector $E\ll U$ of $H(\tau )$. In the limit $%
U\rightarrow \infty $ the Hamiltonian $H(\tau )$ maps to $H_{\mathrm{gauge}}$
with $\lambda _{2}/\lambda _{1}=gf(\tau )/\xi (1-f(\tau ))$. For a critical
value of $\lambda _{1}/\lambda _{2} \sim 0.43$ \cite{numerical,Castelnovo},
the system undergoes a second order QPT.

The lattice gauge theory is dual to the $2+1$-dimensional Ising model \cite%
{Kogut:79}. To show this we define a duality mapping, as follows. To every
plaquette of the original lattice we associate a vertex $p$ on the new
lattice, and we define the first dual variable $\mu ^{x}(p)\equiv B_{p}$. In
order to obtain the correct mapping, the second dual variable must realize
the Pauli algebra with $\mu ^{x}(p)$. Consider now the string $\gamma
_{\rightarrow (\uparrow )}(p)$ from the reference line $t_{2}(t_{1})$ to the
vertex $p$ on the new lattice, see Fig.~\ref{lattice}. To this string we
associate the operator that consists of $\sigma ^{z}$ on every link
intersected by $\gamma _{\rightarrow (\uparrow )}(p)$ and denote it by $\mu
_{\rightarrow (\uparrow )}^{z}(p)$. This is the second dual variable. These
variables realize the Pauli algebra: $(\mu ^{x}(p))^{2}=(\mu ^{z}(p))^{2}=1$
and $\{\mu ^{x}(p^{\prime }),\mu ^{z}(p)\}_{+}=\delta _{p^{\prime }p}$. In
order to write $H_{\mathrm{gauge}}$ in terms of the dual variables, we note
that $\mu _{\uparrow }^{z}(p)\mu _{\uparrow }^{z}(p-\hat{y})=\sigma ^{z}(s,%
\hat{x})$ and $\mu _{\rightarrow }^{z}(p)\mu _{\rightarrow }^{z}(p-\hat{x}%
)=\sigma ^{z}(s,\hat{y})$. Then the mapping $(\sigma ^{x},\sigma
^{z})\mapsto (\mu ^{x},\mu ^{z})$ yields 
\begin{equation}
H_{\mathrm{g}}\mapsto H_{\mathrm{Ising}}=-\lambda _{2}\sum_{p}\mu
^{x}(p)-\lambda _{1}\sum_{p,i}\mu ^{z}(p)\mu ^{z}(p+i),  \label{ising3D}
\end{equation}%
where $\mu ^{z}$ is of the $\rightarrow ,\uparrow $ type if $i=\hat{x},\hat{y%
}$ respectively. We recognize this Hamiltonian as the quantum Hamiltonian
for the $2+1$-dimensional Ising model \cite{Kogut:79}. This model has two
phases with a well understood second order QPT separating them. The gap
between the ground state and the first excited state of this model are known
to scale at the critical point as $\Delta (L)\sim L^{-1}$ where $L={n}^{1/2}$
is the size of the system \cite{Hamer:00}. The first excited state is
non-degenerate. Returning to our $H(\tau )$, and recalling that its low
energy sector $E\ll U$ corresponds to the spectrum of the gauge theory
Hamiltonian (\ref{3Dgauge}), it now follows that also the gap of $H(\tau )$
to the first excited state scales as $\Delta (n)\sim n^{-1/2}$ near the
critical point. Hence we know that for a smooth interpolation the adiabatic
time $T=O(1/\Delta E_{\min })$ scales as $n^{1/2}$, with the final state
arbitrarily close to $|\phi _{0}\rangle $ \cite{Schaller:06}.

\textit{Error estimates}.--- The rest of the spectrum consists of two large
bands with gaps of order $g,U$. We now wish to estimate how well the actual
final state $|\psi (t=T)\rangle $ [the solution of the time-dependent Schr%
\"{o}dinger equation with $H(\tau )$] approximates the desired adiabatic
state $|\phi _{0}\rangle $ at $t=T$ [the instantaneous eigenstate of $H(1)$%
], given that there is \textquotedblleft leakage\textquotedblright\ to the
rest of the spectrum. To this end we use the exponential error estimate \cite%
{hagedorn}: for a continuously differentiable Hamiltonian $H(\tau )$, $%
\delta \equiv \Vert \psi (T)-\phi _{0}(T)\Vert \leq C(n)\exp \left[ -\Delta
(n)T\right] $, where $\Delta (n)$ is the relevant gap and $C(n)\sim \Vert 
\dot{\phi}\Vert /\Delta (n)$, with $\phi (t)$ the instantaneous eigenstate.
Since $g\ll U$, transitions into the lower band dominate. In this case we
can show that $\Vert \dot{\phi}\Vert \sim O(n)$ and hence that already an
adiabatic time as short as $T=(p/g)\log n$ gives an arbitrarily small error $%
\delta \lesssim n^{1-p}$ (for $p>1$) \cite{HLnew}. Therefore the true
adiabatic timescale is set in our case by the previously derived $%
T=O(n^{1/2})$, which completely suppresses errors due to leakage to other
gapped excited states. We have thus shown that the topologically ordered
state $|\phi _{0}\rangle $ can be prepared via adiabatic evolution in a time
that scales as $n^{1/2}$. This is consistent with the strategy devised in 
\cite{Dennis:02}, where a toric code (the ground state $|\phi _{0}\rangle $)
is prepared via a number of syndrome measurements of order $L=n^{1/2}$. This
scaling has recently been shown to be optimal using the Lieb-Robinson bound 
\cite{Bravyi:06Eisert:06}, thus proving that the adiabatic scaling found
here is optimal too.

\textit{Outlook}.--- In this work we have shown how to prepare a
topologically ordered state, the ground state of Kitaev's toric code model,
using quantum adiabatic evolution. The adiabatic evolution time scales as $%
O(n^{1/2})$ where $n$ is the number of spins in the system. This method
gives a measurement-free physical technique to prepare a topologically
ordered state or initialize a topological quantum memory for topological
quantum computation \cite{Freedman:98,Freedman:03,Kitaev:03}.

We can extend Kitaev's model to 3D. In this case, the star operators $A_{s}$
generate surfaces called membranes. It has been shown that this model too
has topological order \cite{Hamma:05}. Its low-energy sector maps onto the
3D quantum lattice gauge theory, whose free-energy expansion shows a \emph{\
first order} QPT between a confined and deconfined (membrane-condensed)
phase \cite{Balian:75}. We therefore conjecture that adiabatic evolution
from a magnetically ordered phase to a membrane-condensed one \cite{Hamma:05}
will involve a first order QPT.

The robustness of the ground state of the Kitaev model to thermal
excitations \cite{Kitaev:03} endows our preparation method with topological
protection against decoherence for $t\geq T$; how to extend this robustness
to all times $t$ is another interesting open question, the study of which
may benefit from recent ideas merging quantum error correcting codes and
dynamical decoupling with adiabatic quantum computation \cite%
{Jordan:05Lidar:07}. An example of the robustness of the topological phase
studied here, when the spin system interacts with an ohmic bath, has
recently been provided in Ref. \cite{Trebst:06}.

Finally, we note that all problems in the computational class
\textquotedblleft Statistical Zero Knowledge\textquotedblright\ (SZK -- a
class which contains many natural problems for quantum computers), can be
reduced to adiabatic quantum state generation \cite{Aharonov:03}, and
preparation of topological order may yield new topological problems in SZK.

\textit{Acknowledgements}.--- We thank S. Haas, A. Kitaev, M. Freedman, S.
Trebst, P. Zanardi and especially M.B. Ruskai and X.-G. Wen for important
discussions. This work was supported in part by ARO W911NF-05-1-0440 and by
NSF CCF-0523675.


\end{document}